\documentclass[sort&compress,
		,final            
  ]
  {aipproc}

\layoutstyle{8x11double}

\begin{document}

\title{Unusual Tunneling Characteristics of Double-quantum-well Heterostructures}

\author{Y.~Lin}{
  address={NTT Basic Research Laboratories, NTT Corporation, Atsugi, Kanagawa 243-0198, Japan}
}

\author{J.~Nitta}{
  address={NTT Basic Research Laboratories, NTT Corporation, Atsugi, Kanagawa 243-0198, Japan}
  ,altaddress={CREST, Japan Science and Technology Agency, Kawaguchi, Saitama 332-0012, Japan}
}

\author{A.~K.~M.~Newaz}{
  address={Department of Physics and Astronomy, SUNY at Stony Brook, Stony Brook, NY 11794-3800, USA}
}

\author{W.~Song}{
	address={Department of Physics and Astronomy, SUNY at Stony Brook, Stony Brook, NY 11794-3800, USA}
}
\author{E.~E.~Mendez}{
	address={Department of Physics and Astronomy, SUNY at Stony Brook, Stony Brook, NY 11794-3800, USA}
}

\begin{abstract}
We report tunneling phenomena in double In$_{0.53}$Ga$_{0.47}$As quantum-well structures that are at odds with the conventional parallel-momentum-conserving picture of tunneling between two-dimensional systems. We found that the tunneling current was mostly determined by the correlation between the emitter and the state in one well, and not by that between those in both wells. Clear magnetic-field-dependent features were first observed before the main resonance, corresponding to tunneling channels into the Landau levels of the well near the emitter. These facts provide evidence of the violation of in-plane momentum conservation in two-dimensional systems.
\end{abstract}

\maketitle

Energy and momentum conservation laws demand that the tunnel current-voltage characteristic of two weakly coupled quantum wells should look like a sharp spike, the current being non-zero only when two states of the wells are aligned in energy.  We have done tunneling experiments in double-quantum-well (DQW) structures that  cast doubts on the validity of such a picture.

The DQW structures we have studied had all the same In$_{0.53}$Ga$_{0.47}$As well thickness design, 53 {\AA} (top) and 82 {\AA} (bottom), but the In$_{0.52}$Al$_{0.48}$As barrier thickness design was different among the structures: all 82 {\AA} (No. 1), all 100 {\AA} (No. 2), and 100 {\AA} with 53 {\AA} in the middle (No. 3). The doping concentration in In$_{0.53}$Ga$_{0.47}$As electrodes was $1 \times 10^{18}$ cm$^{-3}$ outside the well-barrier active region and the 50 {\AA} spacers, and was then increased to $1 \times 10^{19}$ cm$^{-3}$, for about 500 {\AA}. Samples were fabricated into 20$\mu$m$\times 20\mu$m mesas using conventional photolithography and wet etching, and non-alloyed AuGeNi ohmic contacts were formed on the electrodes. At 77K, the I-V characteristics of all three samples exhibited quasi-linear regions followed by distinct negative differential conductance (NDC) features, which changed little as temperature was decreased to 1.8 K.

Figure 1 shows the zero-field I-V characteristics (dotted curves) under positive bias, that is, when electrons tunneled via the wide well into the narrow well.  The current peaks, as shown for Nos. 2 and 3, dovetailed with the energy alignments of two quantum states inside DQWs ($w_{1}$ and $n_{1}$ in the wide and narrow wells, respectively). However, the gradual and broad current increase that preceded each peak corresponded to the passage of those states below the Fermi level of the emitter electrode, apparently violating in-plane momentum conservation. Although such I-V characteristics seem common to many DQW tunneling structures, e.g., Refs. \citealp{Kurata94} and \citealp{Macks96}, to our knowledge this deviation from an in-plane-momentum conserving picture of 2D-2D tunneling has not been emphasized. Furthermore, the tunneling was mainly determined by the correlation between the emitter and one of the wells, not just the far-side well as claimed in Ref.~\citealp{Macks96}. This is clearly observed and supported in the zero-field I-V characteristics for all samples from the onset and the decrease of the current. Based on an analysis of a Poisson-Schr\"odinger calculation, the onset of the current occurred when the Fermi energy in the emitter was aligned with $w_{1}$. The entrance of $n_{1}$ below the Fermi energy did not affect significantly the curvature of the I-V characteristics. When $w_{1}$ and $n_{1}$ passed each other, the current started to decrease up to a certain level, a process still dominated by the resonance of two states. Nevertheless, the current did not drop sharply even after $n_{1}$ passed the conduction-band edge in the emitter, but continued till $w_{1}$ passed below that edge.
\begin{figure}[tp]
\includegraphics[width=.75\columnwidth]{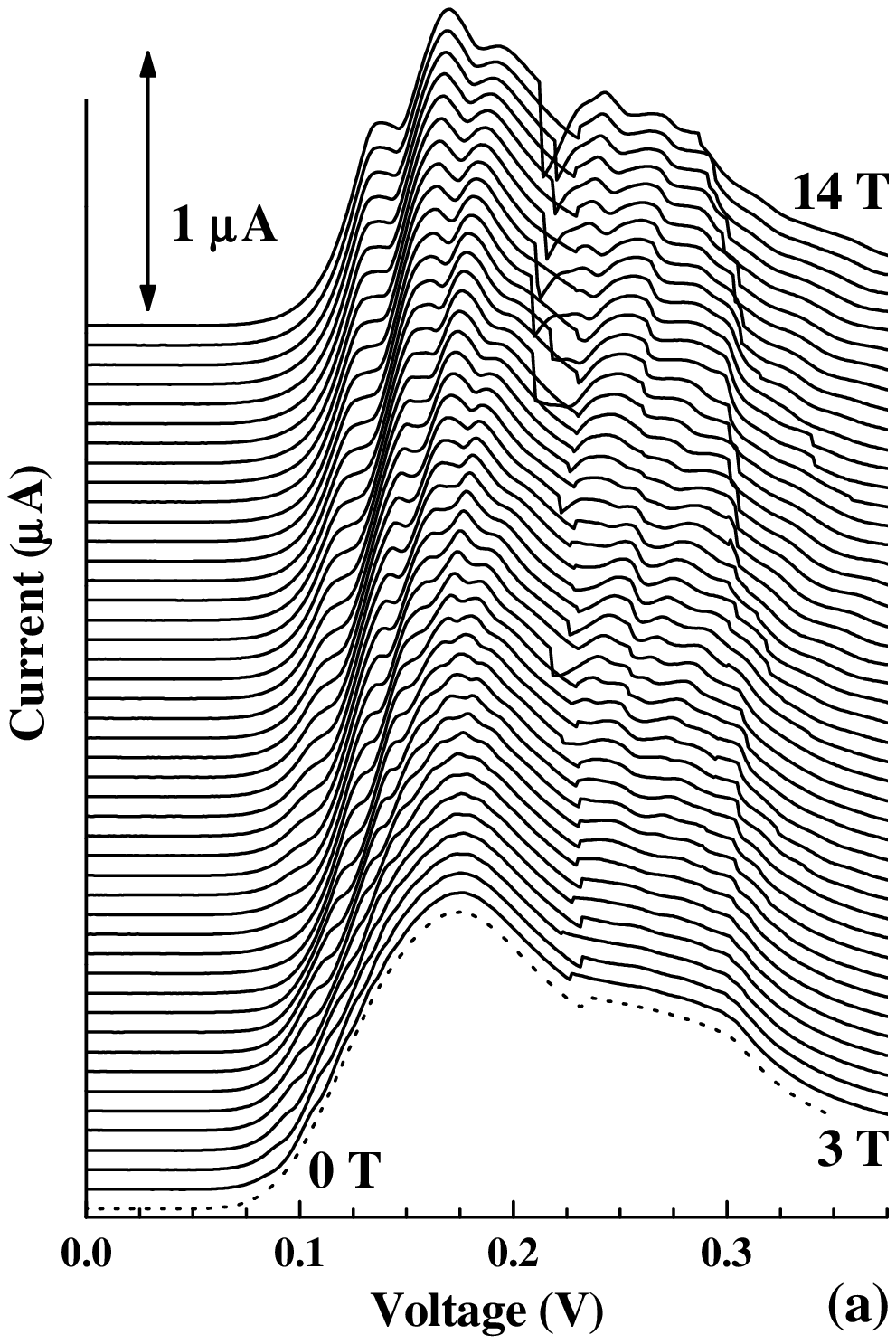}
\includegraphics[width=.75\columnwidth]{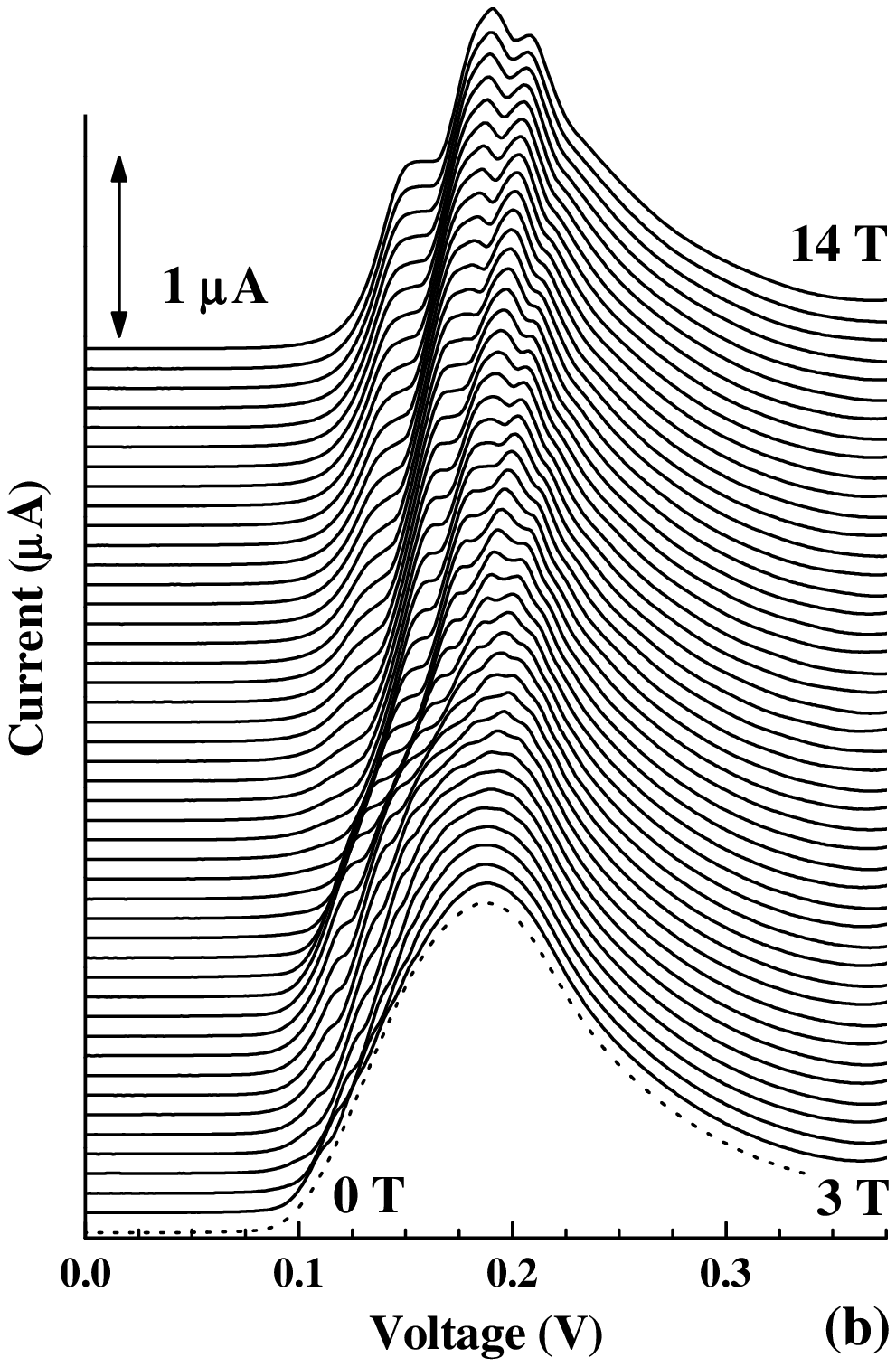}
\caption{Field-dependent I-V characteristics in (a) sample No.~2 and (b) No.~3 from 3 T to 14 T in a step of 0.25 T at 1.8~K. Curves are shifted for clarity. The zero-field I-Vs are plotted with dotted curves. The details are discussed in the main text.}
\end{figure}

Clearer evidence was provided by a magnetic field $B$ ( $\leq$ 14 T) applied parallel to the tunnel current, which revealed new NDC features in the I-V characteristics (see Fig. 1). These features were most prominent in the regions before the current peaks, and moved to higher voltage with increasing field. Such a behavior is incompatible with the usual interpretation of multiple alignments of Landau levels in different wells (even without the conservation of the Landau-level index). On the contrary, it is consistent with a picture in which the field-induced features are related to the passage below the emitter's Fermi level of Landau levels, in the well nearest to the emitter. Ideally 2D-2D tunneling should not show any feature related to Landau levels; even when it did~\cite{Macks96,Asahi91,Wirner97}, the features were until now only seen after the main resonance, via a phonon-assisted process. In contrast, in our samples the features appeared not only before $w_{1}$ and $n_{1}$ were aligned but also before $n_{1}$ was below the emitter's Fermi energy. This surprising fact supports the conclusion drawn from the zero-field I-V characteristics: that before the main resonance the major influence on tunneling comes from the well near the emitter.

Besides these common features in all samples, some details were different from one sample to another. For instance, in Fig.~1 Nos. 2 and 3 differ in the lineshape of the I-V characteristics at $B=0$ and in the nonlinear trends in the motion of Landau levels at certain magnetic fields. The shoulder in NDC [Fig. 1(a)] is usually attributed to phonon-assisted tunneling~\cite{Macks96}, especially when there exists a strong coupling between two states in DQWs. However, this shoulder in the current was seen not on No. 3 but on Nos. 1 and 2, the equal-barrier samples. More pronounced in No. 3 is the sudden nonlinear ``bent'' in the motion of the Landau level between 6 and 9.5 T, after which the Landau level returned to the initial field dependence. This nonlinearity was not absent in other samples; but the field-dependent change was quite small in the lower field range, before 6.5 T. Although the middle barrier in sample No. 3 is rather thin (53 \AA) compared to the external ones (100 \AA), the coupling between two states near the resonance should still be small. It is not clear how these phenomena were affected by a reduction in the thickness of the middle barrier.

To summarize, a study of 2D-2D tunneling has been carried out in In$_{0.52}$Al$_{0.48}$As/In$_{0.53}$Ga$_{0.47}$ double-quantum-well structures. Although the resonance peaks corresponded to the coincidence in energy of two states, the initiation and termination of tunneling were strongly controlled by the well near the emitter. The observation of field-dependent peaks before the main resonance was due to tunneling via Landau levels in that same well. These facts point out the weakness of conservation laws in this system and the unexplained influence of that well on tunneling. Certain field-dependent phenomena are yet to be understood and need further investigation.
%
%
%
%

\end{document}